\documentclass[twocolumn,twoside,slac_two]{revtex4}
\usepackage{graphicx}
\usepackage{fancyhdr}
\usepackage{amsmath,amssymb,amsbsy,amsfonts}
\begin{document}

\title{Hydrodynamical scaling laws for astrophysical jets}

%

\author{Mart\'{\i}n Huarte Espinosa (\textsl{mhuarte@astroscu.unam.mx}) \\
        Sergio Mendoza (\textsl{sergio@astroscu.unam.mx})}
\affiliation{Instituto de Astronom\'{\i}a, Universidad Nacional 
    Autonoma de Mexico, AP 70-264, Ciudad Universitaria,
    Distrito Federal CP 04510, Mexico}
%

\begin{abstract}
   The idea of a unified model for all astrophysical jets has been
considered for some time now. We present here some hydrodynamical scaling
laws relevant for all type of astrophysical jets, analogous to those
of \citet{sams96}.  We use Buckingham's $\Pi$ theorem of dimensional
analysis to obtain a family of dimensional relations among the physical
quantities associated with the jets.
\end{abstract}

\maketitle

\thispagestyle{fancy}


\section{Introduction}

  Although the first report of an astrophysical jet was made by
\citet{curtis}, these objects were extensively studied much later with
radio astronomy techniques \citep{reber}.  Quasars, and radiogalaxies were
discovered and later gathered in a unified model which proposed a dusty
torus around the nucleus of the source \citep{antonucci85}. Years later,
some galactic sources showed similar features to the ones presented by
quasars and radiogalaxies, i.e. relativistic fluxes, a central engine,
symmetrical collimated jets, radiating lobes, and apparent superluminal
motions \citep{sunyaev91}. Optical and X-ray observations showed other
similar non--relativistic sources in the galaxy associated to H-H
objects \citep{pino04}. Lately the strong explosions found in long
Gamma Ray Bursts, had been modelled as collapsars, in which a jet is
associated to the observed phenomena, in order to explain the observations
\citep{kulkarni99,castro99}.

\begin{figure}
  \includegraphics[width=0.45\textwidth]{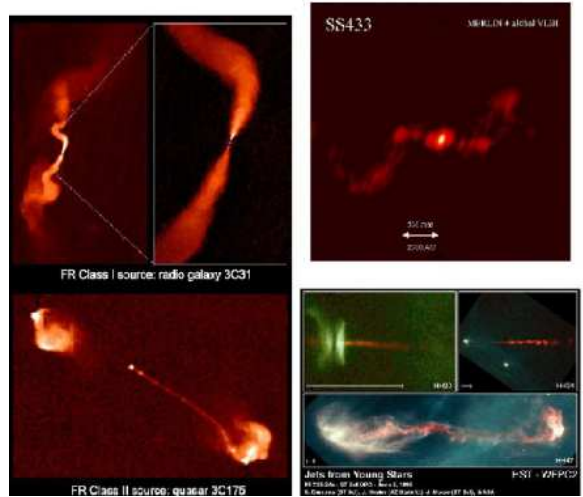}
\caption{ 
     Astrophysical jets are very common and exist in many different
     sizes. On the left, extending $\sim 10^{5} \rm{pc}$, FR~I and FR~2
     sources are shown. The upper right panel shows the micro--quasar
     SS~433. It presents  relativistic fluxes and apparent superluminal
     motions analogous to those in quasars. The lower right panel shows
     jets associated to Herbig--Haro objects with lengths  $\sim 10^{-1}
     - 10 \,\rm{pc}$. All  jets have a condensed (sometimes compact)
     object accreting matter from their surroundings.  There is also
     an accretion disc around the central condensed object and a pair
     of symmetrical collimated jets that end up in radiating lobes.
     The images were taken from \citet{bridle98}, \citet{paragi01} and
     and hubblesite.org.
}
\label{fig1}
\end{figure}

  The similarities between all astrophysical jets, mainly those between
quasars and micro--quasars, and the scaling laws for black holes proposed
by \citet{sams96} and \citet{rees98} made us search for  the possible
existence of some hydrodynamical scaling laws for astrophysical jets.

  The present work presents a few mathematical relations that naturally
appear as a consequence of dimensional analysis and Buckingham's \( \Pi
\) theorem.  We begin by considering some of the most natural physical
dimensional quantities that have to be included in order to describe some
of the physical phenomena related to all classes of jets.  With this
and the use of dimensional analysis we then calculate the dimensional
relations associated to these quantities.  Finally, we briefly discuss
these relations and their physical relevance to astrophysical jets.

\section{Analysis}

  A complete description for the formation of an astrophysical jet
is certainly complicated. However, there are some essential physical
ingredients that must enter into the description of the  problem.
To begin with, the mass \( M \) of the central object must accrete
material from its surroundings at an accretion rate \( \dot{M} \).   Now,
because gravity and magnetic fields \( B \) are necessary in order to
generate jets, Newton's constant of gravity \( G \) and the velocity of
light \( c \)  must be taken into account.  If in addition there is some
characteristic length \( l \) (e.g. the jet's length), a characteristic
density \( \rho \) (e.g. the density of the surrounding medium) and
a characteristic velocity \( v \) (e.g. the jet's ejection velocity),
then the jet's luminosity (or power) \( L \) is a function related to
all these quantities in the following manner

\begin{equation}
  L = L(\dot{M},\, M,\, c,\, G,\,B,\,
  l,\, v,\, \rho). 
 \label{ec4}
\end{equation}

\noindent Using Buckingham's $\Pi$ theorem of dimensional analysis
\citep{buckingham,sedov93} the following non--trivial dimensionless
parameters are found

\begin{alignat}{4}
  \Pi_1 &= \frac{ L }{  \dot{M} c^2 }, & \quad 
  \Pi_2 &= \frac{ G \dot{M} }{ c^3 }, & 
  \Pi_3 &= \frac{ B c^{1/2} M }{ \dot{M}^{ 3/2 } }, \label{pis} \\
  \Pi_4 &= \frac{ l \dot{M} }{ M c },  & \quad 
  \Pi_5 &= \frac{ \rho \ c^3 M^2 }{ \dot{M}^3 }. 
                                                        \notag
\end{alignat}

  From the parameter $\Pi_2$ it follows that  

\begin{equation}   
 \Pi_2 = \left( \frac{G M}{ c^2} \right) \left( \frac{\dot{M}}{
 M} \right) \frac{1}{c}.
\notag 
\end{equation}

\noindent Since the quantity

\begin{equation}
 \tau \equiv \frac{M}{\dot{M}},
 \label{tau}
\end{equation} 

\noindent defines a characteristic time in which the central object
doubles its mass, then using equation~\eqref{tau} we can write \( \Pi_2 \)
as 

\begin{equation}
 \Pi_2 = \frac{r_\text{s}}{ 2 \tau c},
\label{'lambda}
\end{equation}

\noindent where $r_\text{S} $ is the Schwarzschild radius.  This relation
naturally defines a length

\begin{equation}
 \lambda \sim c \tau,
 \label{lambda}
\end{equation} 

\noindent which can be thought of as the maximum possible length a
jet could have, since \( \tau \) is roughly an upper limit to the lifetime
of the source.

  In what follows we will use the
following typical values 

\begin{gather}
  M \approx 10^{8-9} \, \textrm{M}_\odot, \ B \approx 100 \,
    \textrm{G}, \ \dot{M} \approx 1 \textrm{M}_\odot \, \textrm{yr}^{-1}, 
                                                        \notag \\ 
  L \approx 10^{7-10} L_\odot, \ r_\text{j} \approx 10^{4-5} \, \textrm{pc},
                                                        \label{eq01} \\
  \intertext{and}
  M \approx 10^{0-1}  \textrm{M}_\odot, \ B \sim 100 \, \textrm{G}, \ 
    \dot{M} \approx 10^{ -\left( 8-6 \right) } \textrm{M}_\odot 
      \textrm{yr}^{-1},
                                                        \notag \\ 
  L \approx 10^{2 -4} \, L_\odot, \ r_\text{j} 
    \approx 10^{0-1} \textrm{pc},
                                                        \label{eq02}
\end{gather}

\noindent for quasars and \( \mu \)--quasars respectively
\citep{blandford90,carilli96,lovelace96,robson96,reipurth97,ferrari98,shibata98,camenzind99,ford99,meier02,vilhu02,wu02,sunyaev03,smirnov03,calvet04,mendoza04,mirabel04,trimble04}.  

>From equation \eqref{pis} it is found that  

\begin{equation}
 \Pi_6 := \Pi_2^{3/2} \Pi_3 = \left( \frac{ G M
  }{ c^2 } \right)^{3/2}  \bigg/ \quad  \frac{  \sqrt{ M c^2 }  }{B}.   
 \label{ec5}
\end{equation}

\noindent This relation defines a length \( r_\text{j} \) given by

\begin{equation}
  r_\text{j} \propto \frac{ M^{1/3} c^{2/3} }{ B^{2/3} } \approx
    10^2 \left( \frac{ M }{ \textrm{M}_\odot }
    \right)^{1/3}  \left( \frac{ B }{ 1 \rm{G} } \right)^{-2/3}
    \rm{pc}. 
\label{r_j}
\end{equation}

 For typical extragalactic radio sources and \( \mu \)--quasars 
it follows from equations~\eqref{eq01} and \eqref{eq02} that $r_\text{j}
\propto 10^{4} \, \rm{pc}$ and  $r_\text{j} \propto 10
\, \rm{pc}$ respectively.  These lengths are fairly similar to the
associated length of these jets.  In other words, if we identify the
length \( r_\text{j} \) as the length of the jet, then a constant of
proportionality $ \sim 1 $ is needed in equation~\eqref{r_j}, and so

\begin{equation}
 r_\text{j} \approx 100 \left( \frac{ M }{ \textrm{M}_\odot } \right)^{1/3}  
 \left( \frac{ B }{ 1 \rm{G} } \right)^{-2/3} \rm{pc}. 
 \label{r_j=}
\end{equation}

Since equation~\eqref{ec5} is roughly the ratio of the Schwarzschild
radius $r_\text{S}$ to the jet's length $r_\text{j}$, then $\Pi_6 << 1$, i.e. 

\begin{equation} 
 \Pi_6 = \frac{ \left( B l^{3/2} \right) \left( G M^2/l
 \right)^{3/2} }{ (M c^2 )^2} << 1,
\label{r_j''}
\end{equation}
\noindent which in turn implies that
\begin{equation} 
 B << \frac{ c^4 }{ G^{3/2} M } \approx 10^{23} \left(
 M/\textrm{M}_\odot \right)^{-1} \, \rm{G}.
\label{r_j'2}
\end{equation}

The right hand side of this inequality is the maximum upper limit for
the magnetic field associated to the accretion disc around the central
object. For ``extreme'' micro--quasars like SS~433 and GRB's the magnetic
field $ B \gtrsim 10^{16} \, \rm{G}$, so that this upper limit works
better for those objects \citep{meier02,trimble04}.

   From equation~\eqref{pis} it follows that

\begin{equation}   
 \Pi_7 \equiv \frac{\Pi_1}{\Pi_2 \Pi_3^2} = \frac{L \dot{M}}{
 B^2 M^2 G },
\notag 
\end{equation}   

\noindent and so

\begin{equation}   
 L \propto 10^{-7} \left( \frac{B}{1 \, \rm{G}} \right)^2  
 \left( \frac{M}{\textrm{M}_\odot} \right)^2  \left(
 \frac{\dot{M}}{\textrm{M}_\odot yr^{-1}}
 \right)^{-1} L_\odot.
\label{ls}
\end{equation}   

For the case of quasars and \( \mu \)--quasars, using the typical values of
equations~\eqref{eq01} and \eqref{eq02} it follows that 
the power $L \propto 10^{15} \, L_\odot$ and $L \propto  10^{8} \, 
L_\odot$ respectively.  In order to normalise it to the observed values, we
can set a constant of proportionality $\sim 10^{-6} $ in
equation~\eqref{ls}.  With this, the jet power relation is given by

\begin{equation}   
  L \approx 10^{-13} \left( \frac{B}{ 1 \, \rm{G} } \right)^2  
  \left( \frac{M}{\textrm{M}_\odot} \right)^2  \left(
  \frac{\dot{M}}{\textrm{M}_\odot yr^{-1}}
  \right)^{-1} L_\odot.
  \label{ls'}
\end{equation}

\section{Conclusion}

  Astrophysical jets exist due to a precise combination of
electromagnetic, mechanic and gravitational processes, independently of
the nature and mass of their central objects. 
  
  Here we report the dimensional relation between a few important
parameters that enter into the description of the formation of an
astrophysical jet.  

  Of all our results, it is striking the fact that the jet power is
inversely proportional to the accretion rate associated with it.  This
is probably due to the following.  For a fixed value of the
mass of the central object (in any case, for the time that accretion
takes place, the mass of the central object does not increase too much)
when the accretion mass rate increases, then the magnetic field lines
anchored to the plasma tend to pack up, meaning that the field intensity
increases in such a way as to get the correct result given by
equation~\eqref{ls'}.

\section{Acknowledgements}

  We would like to thank S. Setiawan for useful discussions about jet
power and their association with gravitational effects.  SM gratefully
acknowledges support from DGAPA (IN119203) at Universidad Nacional
Aut\'onoma de M\'exico (UNAM).

\bibliography{poster-proc}





\end{document}